\documentclass[sigconf]{acmart}

\AtBeginDocument{%
  }

\setcopyright{acmlicensed}
\copyrightyear{2025}
\acmYear{2025}
\setcopyright{rightsretained}
\acmConference[UIST Adjunct '25]{The 38th Annual ACM Symposium on User Interface Software and Technology}{September 28-October 1, 2025}{Busan, Republic of Korea}
\acmBooktitle{The 38th Annual ACM Symposium on User Interface Software and Technology (UIST Adjunct '25), September 28-October 1, 2025, Busan, Republic of Korea}
\acmDOI{10.1145/3746058.3758427}
\acmISBN{979-8-4007-2036-9/2025/09}

\begin{document}

\title[RoboLinker]{RoboLinker: A Diffusion-model-based Matching Clothing Generator Between Humans and Companion Robots}

\author{Jing Tang}
\affiliation{
  \institution{Huazhong University of Science and Technology}
  \city{Wuhan}
  \country{China}
}
\email{j_tang@hust.edu.cn}

\author{Qing Xiao}
\affiliation{
  \institution{Carnegie Mellon University}
  \city{Pittsburgh}
  \state{Pennsylvania}
  \country{USA}
}
\email{qingx@andrew.cmu.edu}

\author{Kunxu Du}
\affiliation{
  \institution{Future Design School, Harbin Institute of Technology, Shenzhen}
  \city{Shenzhen}
  \country{China}
}
\email{dux2902996813@gmail.com}

\author{Zaiqiao Ye*}
\affiliation{
  \institution{Future Design School, Harbin Institute of Technology, Shenzhen}
  \city{Shenzhen}
  \country{China}
}
\email{yezaiqiao@hit.edu.cn}

\begin{abstract}
We present RoboLinker, a generative design system that creates matching outfits for humans and their robots. Using a diffusion-based model, the system takes a robot image and a style prompt from users as input, and outputs a human outfit that visually complements the robot's attire. Through an interactive interface, users can refine the generated designs. We evaluate RoboLinker with both humanoid and pet-like robots, demonstrating its capacity to produce stylistically coherent and emotionally resonant results. 
\end{abstract}

\begin{CCSXML}
<ccs2012>
   <concept>
       <concept_id>10010583.10010682.10010689</concept_id>
       <concept_desc>Computer systems organization~Robotics</concept_desc>
       <concept_significance>500</concept_significance>
   </concept>
</ccs2012>
\end{CCSXML}

\ccsdesc[500]{Computer systems organization~Robotics}

\keywords{Human-Robot Interaction, Matching Outfits, Robot Fashion}

% make the title area
\maketitle

\section{Introduction}
Emotional attachment between humans and robots is increasingly seen as a key challenge in Human-Robot Interaction (HRI) \cite{walters2008design, konok2018should,dautenhahn2004robots,chen20253r,chen2025sustainable}. Among various strategies to foster this bond, clothing has emerged as a compelling yet understudied medium \cite{Designing2021Friedman}. Recent trends suggest robot fashion is evolving beyond aesthetics into a form of social interface. Researchers have explored how clothing enhances robot presence, functionality, and user expression \cite{joshi2024social}. For instance, Ye et al. \cite{Dressing2023Ye} show users dress AIBO robots to express personality, seek comfort, or join creative communities. Despite this momentum, most work focuses only on the robot as a display object. What remains underexplored is fashion as a two-way medium: not only styling robots, but allowing humans to express their emotional ties through what they themselves wear. Just as matching outfits signal closeness among couples or families, we propose that coordinated clothing in HRI can make human-robot relationships visible and expressive. 

To address this, we present \textit{RoboLinker}, a novel digital tool that enables humans and companion robots to co-design matching outfits using a diffusion model approach. By allowing users to input an image of their robot, along with a descriptive prompt for a desired style, RoboLinker generates \textit{a personalized outfit for the human owner that visually complements the robot’s look}.

\begin{figure}[t]
  \centering
  \includegraphics[width=0.9\linewidth, trim=20 20 20 20, clip]{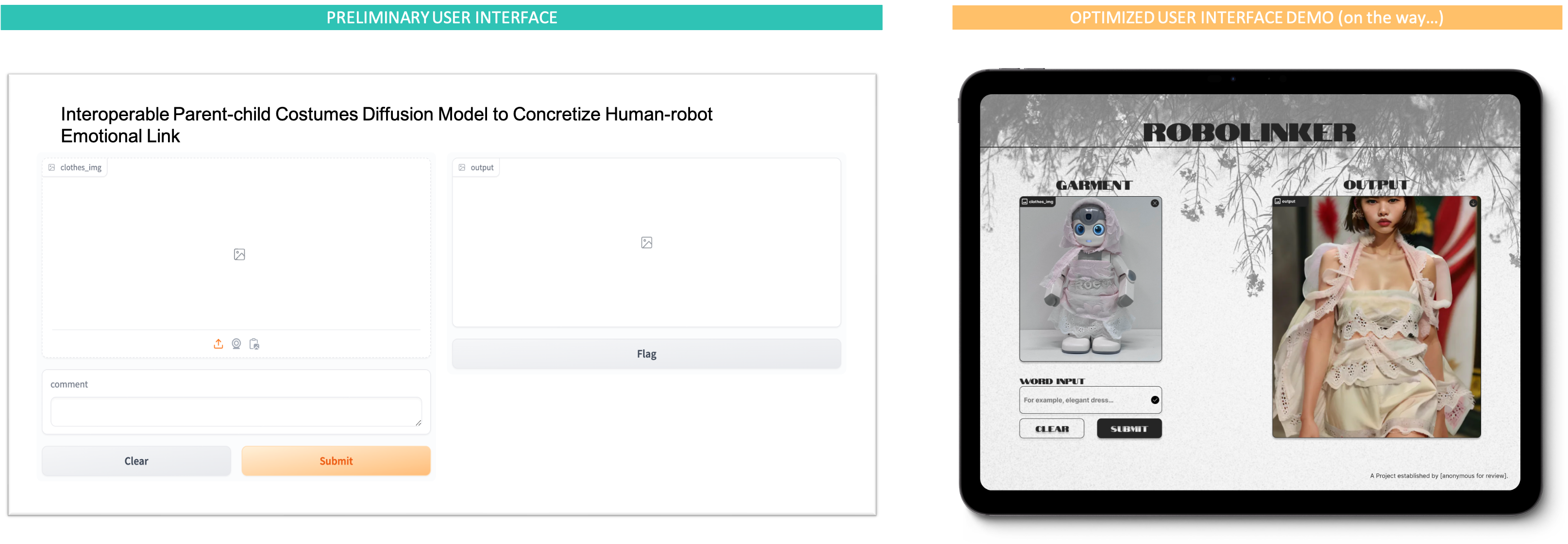}
  \caption{The RoboLinker interface. Users upload a robot image and a style prompt, and receive a generated human outfit that complements the robot’s look.}
  \label{fig2}
\end{figure}

\section{RoboLinker}
\subsection{Technical Design}
We simulate a collaborative outfit design approach between robots and humans. We use the language of humans and the outfits of robots as inputs to the human and robot sides respectively. These are fed into a diffusion model named RoboLinker designed for this task to enable communication between ``human buddy" (Human Preference of Clothing Style by Language) and ``robot buddy" (Robot Clothing) in the process of corresponding human outfit design. We first make a unified mathematical representation of opinions between ``human buddy" and ``robot buddy". Therefore, here we consider tokenization of the text input by the ``human buddy" and the image input by the ``robot buddy" to obtain uniform mathematical vectors $(Latent_1, Latent_2, ... Latent_N)$. In the latent space of tokens, the ``human buddy" and ``robot buddy" can create different digital representations for different entities in the real world using the tokenization technique. As for the intense discussion between the ``human buddy" and the ``robot buddy", it can be realized by continually tokenizing their inputs and feeding the tokens into the diffusion model to realize the gradual optimization of the final corresponding ``human buddy" outfit. These processes can be formulated as:

\begin{equation}
    Token_{human buddy} = Tokenization(Text_{human buddy})
\end{equation}
\begin{equation}
    Token_{robot buddy} = Tokenization(Image_{robot buddy})
\end{equation}
\begin{equation}
    Clothes = RoboLinker(Token_{human buddy},Token_{robot buddy})
\end{equation}

\begin{figure}[t]
  \centering
  \includegraphics[width=0.95\linewidth, trim=10 10 10 10, clip]{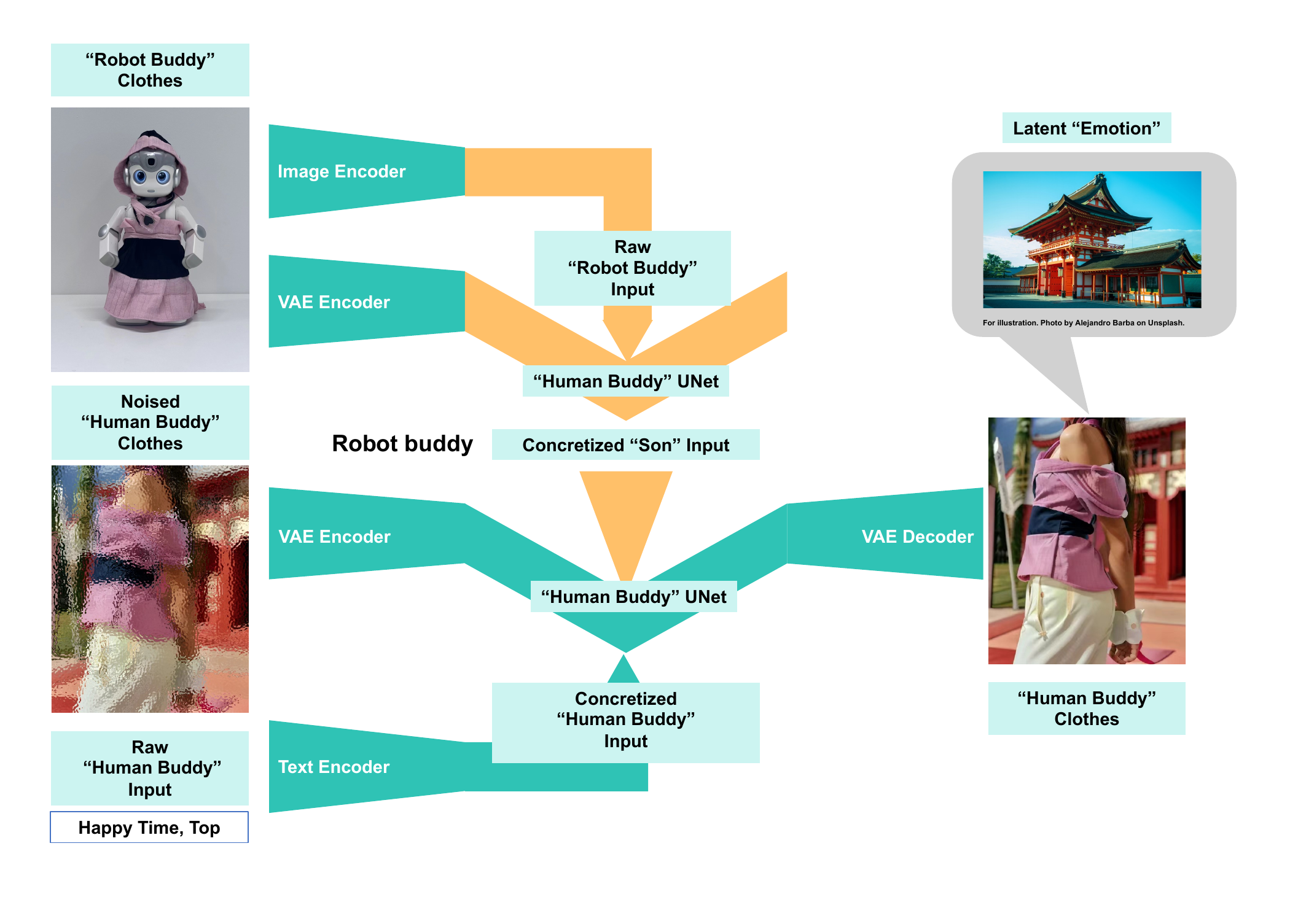}
  \caption{Model architecture of RoboLinker. The system encodes robot and human fashion inputs and generates coordinated human outfits via a diffusion-based pipeline.}
  \label{fig1}
\end{figure}

The above unified token representation is equivalent to inputting the views of ``human buddy" and ``robot buddy" into a unified latent space. We need to build a diffusion model to understand the unified representation. At the same time, this diffusion model needs to support the ``human buddy" and the ``robot buddy" to co-create and ensure that the final design result can meet the needs of the ``human buddy" and the ``robot buddy". In addition, we hope that the diffusion model designed can stimulate some potential empathy points of the ``human buddy" and ``robot buddy" to realize a deeper emotional matching outfits design. The detailed model architecture of RoboLinker is shown in \autoref{fig1}.

\subsection{Interactive Interface}

The interactive interface for the ``RoboLinker" is shown in Figure \ref{fig2}. It includes two parts: (1) \textit{Model Image Upload and Initial Design}: Users initiate the process by uploading an image of a robot wearing a garment. Upon receiving the image, the system utilizes advanced image recognition algorithms to analyze key elements such as the garment's style, color, fabric type, and specific details like embellishments or stitching. Based on this analysis, the system generates an initial textual description of the garment, highlighting aspects such as silhouette, fabric texture, and design features. Users are then provided with multiple adjustment options with initial human clothing design outputs, which are tailored to the garment’s extracted attributes. (2) \textit{Iterative Design}: Once the initial design description is generated, users can provide feedback on elements they wish to modify, such as the color, fabric, or cut of the garment, and write their preferences. The system then processes this feedback in real time, adjusting the design accordingly. For instance, users may request alterations to the garment’s fit, such as modifying the hemline. The iterative design cycle continues until the user is satisfied with the modifications.

\begin{figure}[t]
  \centering
  \includegraphics[width=0.9\linewidth, height=0.4\textheight, keepaspectratio]{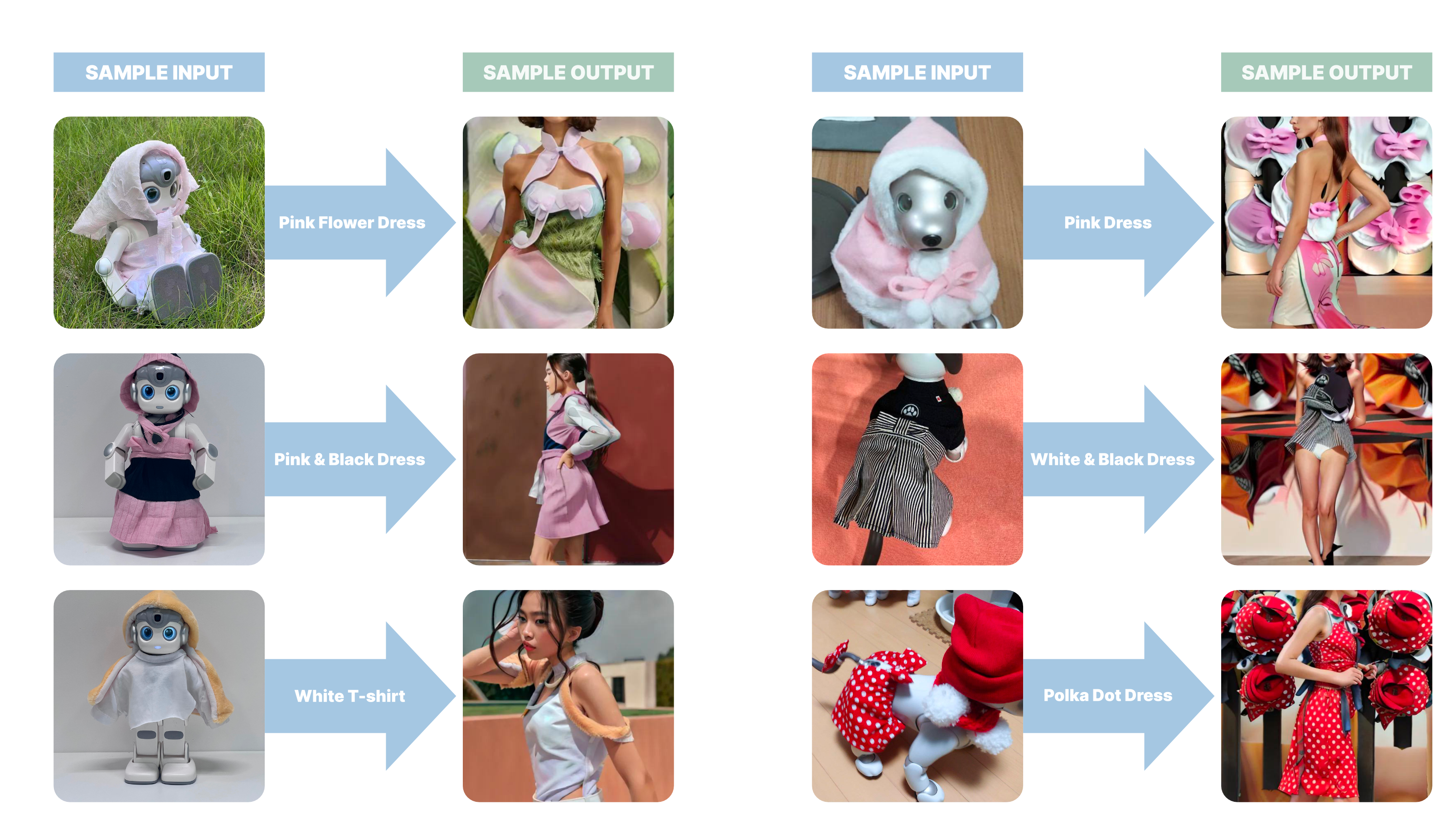}
  \caption{Sampled outcomes of RoboLinker. Each column shows robot images and generated human outfits.}
  \label{fig3}
\end{figure}

\section{Technical Evaluation}
To test the effect of RoboLinker, we selected two different inputs including wukong-robot and AIBO robot dog, which represent humanoid companion robot input and pet-like companion robot input. Our technical evaluations are presented in \autoref{fig3}, with results for the Wukong robot shown on the left and those for the AIBO robot dog on the right.

\textbf{Wukong-robot with Clothes}: Wukong-robot is a popular humanoid Chinese voice dialogue robot project ~\cite{wukong-robot}. We designed different styles of clothes for the wukong-robot, collected image data based on our design, and input them into the model together with some expectations of the ``human buddy" for the preliminary test. In this example, we designed denim-style clothes for the wukong-robot, and the resulting matching outfits also inherited this style, which shows the superiority of the model.

\textbf{AIBO Robot Dogs with Clothes}: AIBO is a series of robotic dogs designed and manufactured by Sony \cite{Dressing2023Ye}. We collected a large amount of AIBO robot dogs with clothes images from user posts on X (formerly Twitter) to test our model. It can be seen that the clothes of the ``human buddy" generated in the end contain all the elements of AIBO, which indicates that our model can indeed understand the meaning of ``human buddy" and ``robot buddy" very well, and achieve a very stable output of matching outfits.

\section{Conclusion}
RoboLinker allows users to generate and customize human clothing that visually and stylistically matches their robots, using a diffusion-based generative model and a human-robot co-design pipeline. By enabling users to co-create matching clothing, RoboLinker helps turn the act of dressing into a form of relational authorship between human and robot. Future directions include enhancing its ability to adapt to diverse robot forms and postures, as well as supporting reverse generation from human clothing to robot apparel.

\bibliographystyle{ACM-Reference-Format}
\bibliography{reference}

\appendix

\end{document}